\documentclass[twocolumn,preprintnumbers,amssymb,amsmath,superscriptaddress,letterpaper]{revtex4}[10pt]

\usepackage{graphicx}
\usepackage{dcolumn}
\usepackage{bm}
\usepackage{natbib}
\usepackage{pstricks}

\newcommand{\be}{\begin{equation}}
\newcommand{\ee}{\end{equation}}
\newcommand{\ba}{\begin{eqnarray}}
\newcommand{\ea}{\end{eqnarray}}

\newcommand{\GHz}{{\rm ~GHz}}

\newcommand{\GeV}{{\rm ~GeV}}

\begin{document}

\title{An Alternative Approach for Constraining the Galactic Magnetic Field and the Interstellar Radiation Field}

\author{Maryam Tavakoli}
\email{maryam.tavakoli@ipm.ir}
\affiliation{School of Astronomy, Institute for Research in Fundamental Sciences (IPM), P.O. Box 19395-5531, Tehran, Iran}

\date{\today}

\begin{abstract}
High energy cosmic ray electrons and positrons, during propagation in the Galaxy, mainly lose energy through either the inverse Compton scattering (ICS) off of the interstellar radiation field (ISRF) or emitting synchrotron radiation in the Galactic magnetic field (GMF). Emitted photons by these two main energy loss mechanisms contribute to, respectively, diffuse Galactic gamma rays and radio waves. The variation of the distribution of the ISRF and the GMF has a sizable impact on the spectrum of cosmic ray electrons and positrons throughout the Galaxy. This impact is revealed both locally through measuring the local flux of cosmic ray electrons and positrons and globally by observing the spectra and morphology of diffuse gamma rays and radio waves. In this paper, we quantify this impact and conclude that recent data from AMS-02, \textit{Fermi} and H.E.S.S experiments are powerful enough to be alternatively applied for constraining properties of the ISRF and the GMF.
\end{abstract}

\maketitle


\section{Introduction}
\label{sec:intro}

The GMF and the ISRF are two interstellar medium ingredients whose energy densities are the same order to that of cosmic rays. Thus, they play an essential role in determining the dynamics and processes in the interstellar medium. 

The GMF is inferred from the Faraday rotation measure and the synchrotron radiation in total intensity and polarization \cite{Jansson:2009ip,2012ApJ...761L..11J, 2012ApJ...757...14J,han17}.
It has a regular component which is believed to follow the spiral structure of the Galaxy with a pitch angle $p$.
The direction of the local regular magnetic field is defined as $l=90^{\circ}+p$ where $l$ the Galactic longitude points to the Galactic center when $l=0^{\circ}$ and points to the south when $l=90^{\circ}$.
There is no consensus on the exact morphology of the regular magnetic field. Indeed, a variety of models for the ordered component of the GMF can be found in the literature (see \cite{2016A&A...596A.103P} and references therein).
The GMF has also a turbulent component.  
Because of the random direction, the turbulent magnetic field only contributes to the total intensity of the synchrotron radiation without any effect on either the Faraday rotation measure or the polarized synchrotron emission.
The turbulent magnetic field intensity is conventionally assumed to exponentially decrease outward the Galaxy in both radial and vertical directions with the maximum value at the Galactic center
\be
B_{turb}=B_{0,turb}\exp{(-\frac{R-R_{\odot}}{R_{0,turb}})}\exp{(-\frac{|z|}{z_{0,turb}})}.
\label{eq:Bturb}
\ee
The turbulent component of the GMF is much less certain than the ordered one. Studies on polarization suggest the local energy density of turbulent magnetic field is roughly equal to that of regular magnetic field \cite{1971A&A....14..359B, 1976A&AS...26..129B, 1981A&A....98..286P, 1992ApJ...389..602J, 2008A&A...490.1093M}.
For a given value of $B_{0,turb}$, the intensity of the magnetic field at the Galactic centre $B_{GC}$, which is dominated by the turbulent component, is dictated by the radial scale of the turbulent magnetic field $R_{0,turb}$.
The Galactic centre magnetic field intensity is largely uncertain. There is a lower bound of $50\mu$G \cite{2010Natur.463...65C} and the most probable value is $100\mu$G.
The scale height $z_{0,turb}$ is assumed to be the same as the scale height of the diffusion zone. 

The ISRF stands for low energy photons in our Galaxy. It consists of optical, infrared and cosmic microwave background (CMB) photons.
Optical photons are emitted by stars and infrared photons are scattered or absorbed star light by dusts. 
The distribution and the spectrum of optical and infrared photons are calculated by using the stellar and dust distribution taking the absorption and scattering processes into account \cite{2005ICRC....4...77P}. 
The CMB energy density is well known by direct observations \cite{2016A&A...594A...9P}. 

The GMF and the ISRF can be further constrained noting that high energy cosmic ray electrons and positrons mainly lose energy by ICS off of the ISRF and by synchrotron emission in the GMF \cite{2017JCAP...05..001B}.
The morphology and the energy density of the GMF and the ISRF affect the energy loss of high energy cosmic ray $e^{\pm}$s and thus affect their spectrum in the Galaxy. 
This influence can be locally observed by measuring the local spectrum of cosmic ray electrons and positrons.
Redistributed cosmic ray electrons and positrons produce photons via ICS and synchrotron emission with different morphology and spectrum.
This global effect can be observed in spectra of diffuse gamma rays and radio waves.
Thus, the distribution of the ISRF indirectly impacts the spectrum of the synchrotron emission and so it is for indirect impact of the distribution of the GMF on the spectrum of ICS component of diffuse gamma rays. 
In this paper, we quantify those impacts. 
Indeed, high accuracy data for the spectra of cosmic ray $e^{\pm}$s by AMS-02 \cite{PhysRevLett.113.121101, 2014PhRvL.113l1102A, PhysRevLett.113.221102, Abdollahi:2017nat} and for the spectra of diffuse gamma rays by \textit{Fermi}-LAT \footnote{http://fermi.gsfc.nasa.gov/ssc/data/analysis/scitools/} and H.E.S.S \footnote{https://www.mpi-hd.mpg.de/hfm/HESS/} have the power to put constraints on properties of the GMF and the ISRF.  
In section \ref{LISM}, we show how the local spectra of cosmic ray electrons and positrons change by varying the properties of the local interstellar medium. Section \ref{GISM} is devoted to effects of global properties of the GMF and the ISRF on the spectra of ICS component of diffuse gamma rays and synchrotron component of radio waves. Our conclusions are presented in \ref{Conclusion}.

\section{Local Interstellar Medium}
\label{LISM}

The flux of cosmic ray $e^{\pm}$s is composed of primary electrons, which are accelerated by supernova remnants, secondary electrons and positrons, which are produced by interaction of cosmic ray protons and heavier nuclei with the interstellar gas, and high energy electrons and positrons which are injected by extra sources such as pulsars and/or dark matter \cite{2014PhRvL.113l1102A,PhysRevLett.113.221102,PhysRevLett.113.121101}. 
Cosmic ray electrons and positrons, after being accelerated by sources, are injected into the interstellar medium where they diffuse and lose energy by a  number of mechanisms. 
The estimate of the flux at low energies has large uncertainties. 
The secondary production cross section below $E=10\GeV$ is not known well. 
Also, at low energies the spectrum of secondaries is sensitive to the diffusion scale height. 
A thinner diffusion zone predicts a smaller diffusion coefficient. This relies on the local spectrum of Boron to Carbon ratio in agreement with AMS-02 data \cite{PhysRevLett.117.231102}.
The smaller diffusion coefficient makes cosmic ray protons and heavier nuclei stay longer close to the Galactic disk, where the interstellar gas is denser, and produce more secondaries.
Moreover, the spectrum at energies below $E=10\GeV$ is time and charge dependent because of modulation by the solar activity \cite{2014PhRvL.113l1102A,Cholis:2015gna}. 
To avoid those uncertainties, we perform our analysis using data with energy above $10\GeV$.
In high energies, ICS off of the ISRF and synchrotron emission in the GMF are dominant energy loss processes \cite{2017JCAP...05..001B}. 
The rates of energy loss by those processes sharply increase with the energy of cosmic ray electrons and positrons. Thus, high energy cosmic ray $e^{\pm}$s that reach us have only probed the local interstellar medium.
The energy loss rates via ICS and synchrotron emission are proportional to, respectively, the ISRF and the GMF energy densities 
\be
\Big(\frac{dE}{dt}\Big)_{ICS} = \frac{4}{3}\sigma_Tcu_{ISRF}\Big(\frac{v}{c}\Big)^2\gamma^2  \nonumber 
\ee
\be
\Big(\frac{dE}{dt}\Big)_{sync} = \frac{4}{3}\sigma_Tcu_{GMF}\Big(\frac{v}{c}\Big)^2\gamma^2
\ee
where $\sigma_T$ is the Thomson scattering cross section, $u_{ISRF}$ is the energy density of radiation, $u_{GMF}$ is the energy density of the magnetic filed, $v$ is the electron velocity and $\gamma$ is its Lorentz factor.
The ICS energy loss rate with $\sigma_T$ is valid so long as $\gamma \hbar \omega \ll m_ec^2$. For ultra relativistic electrons, $\gamma \gg 1$, the Thomson cross section must be replaced by the Klein-Nishina cross section $\sigma_{K-N}$ which is inversely proportional to the energy of photons. 
Thus, in high energies the ICS off of CMB photons is more effective than that of IR and optical photons.
The amount of energy that electrons lose by ICS is $(4/3 \gamma^2-1)E_0$ where $E_0$ is the initial energy of photons. A higher energy electron loses more energy while up-scattering a photon compared to a lower energy electron.
On the other hand, the synchrotron emission becomes maximum at $\nu_{max}=0.29\frac{3}{2}\gamma^2\nu_g\sin\alpha$
where $\nu_g=eB/2 \pi m_e=28 B \GHz T^{-1}$ and $\alpha$ is the pitch angle of electrons orbit.
Cosmic ray electrons with higher energy emit synchrotron radiation at higher frequencies. 
In a nutshell, the higher energy cosmic ray $e^{\pm}$s, the higher rate and the higher amount of energy loss. 

\begin{figure}
\hspace{-0.5cm}
\includegraphics[scale=0.4]{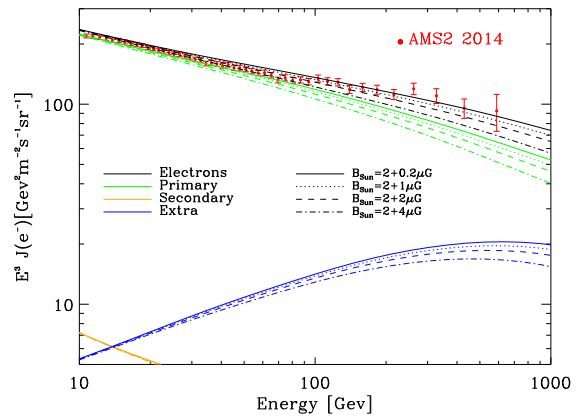}
\caption{Spectrum of cosmic ray electrons for various local turbulence intensities. The local turbulent magnetic field is $0.1$ (\textit{solid line}), $0.5$ (\textit{dotted line}), $1$ (\textit{dashed line}) and $2$ (\textit{dashed-dotted line}) times the local regular magnetic field which is $2\mu$G.}
\label{fig:e_multi_Aturb}
\end{figure}

Here, we investigate how the local flux of cosmic ray electrons changes by varying the ISRF and the GMF properties.
First, we examine different GMF morphologies. We choose three representative models for the regular magnetic field: the bi-symmetric spiral \cite{2008A&A...490.1093M}, the axisymmetric spiral with rings \cite{2008A&A...477..573S} and the logarithmic spiral \cite{2011ApJ...738..192P}. 
As long as the local magnetic field intensities are the same, different GMF morphologies can not be discriminated by the local flux of cosmic ray electrons.  
The relative difference of the flux for those GMF morphologies is less than $2\%$. 
The local fluxes of cosmic ray electrons and positrons are expected to be mostly sensitive to the local interstellar medium properties.
In Fig.~\ref{fig:e_multi_Aturb}, we show how the intensity of the local turbulence affects the flux of cosmic ray electrons. We vary the local turbulent magnetic field from tenth to twice the local regular magnetic field that is $B_{\odot,regular}=2\mu$G. 
As expected, by increasing the local magnetic field intensity, the energy loss rate increases and the flux falls more steeply.

\begin{figure}
\hspace{-0.5cm}
\includegraphics[scale=0.4]{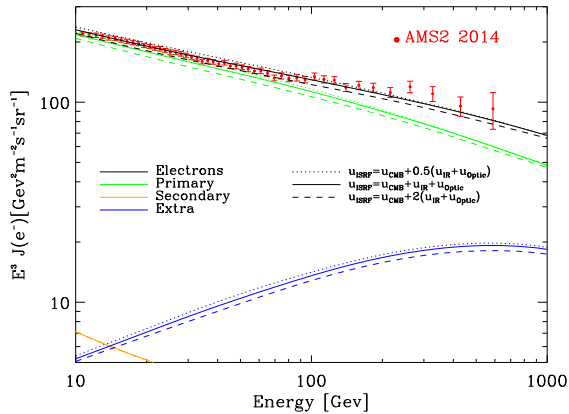}
\caption{Local flux of cosmic ray electrons for different ISRF energy densities. Having fixed the CMB energy density, we vary the energy density of infrared and optical components by multiplying them by $0.5$ (\textit{dotted line}) and $2$ (\textit{dashed line}). The reference ISRF energy density is shown as \textit{solid line}.}
\label{fig:e_multi_ISRF}
\end{figure}

In Fig.~\ref{fig:e_multi_ISRF}, the impact of varying the ISRF energy density on the spectrum of cosmic ray electrons is shown. 
Since the CMB energy density is certain from observations, we keep it fixed and only multiply the energy density of optical and infrared photons by a factor. 
By increasing the ISRF energy density, the rate of energy loss through ICS increases and the spectrum becomes softer.
However, we note that softening of the flux in high energies is fainter. The reason is that in high energies the Thomson scattering cross section is not valid any longer and it must be replaced by $\sigma_{K-N}$. Thus, the increase in the energy loss rate in high energies is less abrupt. 

In addition to energy loss processes, the properties of primary sources of cosmic ray $e^{\pm}$s and the properties of cosmic ray diffusion affect the spectrum. 
The leaky box approximation predicts that the energy spectrum of primary cosmic ray electrons is proportional to $E^{-(\gamma_e+\delta/2)}\tau^{1/2}_{loss}$ where $\gamma_e$ is the injection spectral index of primaries, $\delta$ is the diffusion spectral index and $\tau_{loss}$ is the time scale of the energy loss.
A full analysis of all parameters, which are involved in determining the local spectra of cosmic ray electrons and positrons, has been performed in \cite{Maryam}. 


\section{Global Interstellar Medium}
\label{GISM}

\begin{figure}[htb]
\hspace{-0.5cm}
\includegraphics[scale=0.4]{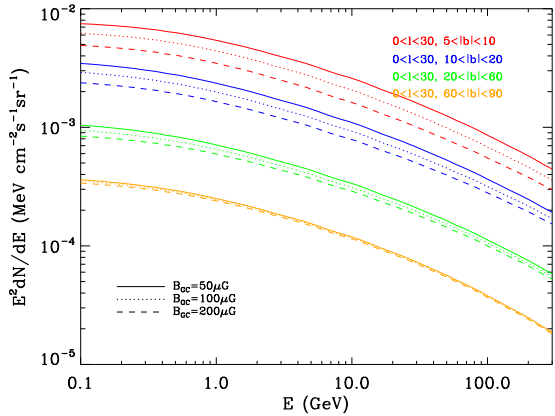}
\caption{Spectrum of ICS in different sky windows for different magnetic field intensities at the Galactic centre. The value of $B_{GC}$ is $50\mu$G (\textit{solid line}), $100\mu$G (\textit{dotted line}) and $200\mu$G (\textit{dashed line}).}
\label{fig:ICS_multi_BGC}
\end{figure}

Diffuse gamma rays are mostly produced in our own Galaxy.
Decay of neutral pions which are produced via interaction of cosmic ray nuclei with the interstellar gas, ICS of high energy $e^{\pm}$s off of the ISRF and bremsstrahlung of high energy electrons and positrons in the interstellar gas give rise to gamma rays. 
Our Galaxy is transparent to GeV scale gamma rays. Thus, those photons propagate on direct lines through the Galaxy without any attenuation.
Gamma rays that reach us from a given direction carry information about properties of the interstellar medium along that line of sight.
As for up scattered photons via Compton phenomenon, the energy density of the ISRF in different sky regions directly affects the ICS spectrum. The more ISRF energy density, the harder ICS spectrum. 
Moreover, assumptions on properties of the GMF have indirect impacts on ICS spectra by altering the dominant mechanism of energy loss of high energy cosmic ray $e^{\pm}$s. The stronger GMF, the more energy loss via synchrotron emission, the softer ICS spectrum.
In order to quantify those effects, we first implement different morphologies for the regular magnetic field as mentioned in Sec. \ref{LISM}. 
The budget of energy density in regular magnetic field is almost the same for various morphologies. Besides, in most sky regions the turbulent magnetic field dominates the energy density. 
Thus, various morphologies slightly affect the rate of energy loss of cosmic ray $e^{\pm}$s and leave the ICS spectra almost unchanged \footnote{The ICS spectra change by $5\%$ at most.}.
We choose the logarithmic spiral model \cite{2011ApJ...738..192P} for the regular magnetic field and only vary the turbulent magnetic field. 
The local turbulence is constrained by local spectra of cosmic ray electrons and positrons. 
We fix $B_{0,turb}$ and increase the intensity of the GMF at the Galactic centre by decreasing the radial scale of turbulence $R_{0,turb}$. 
The increase of $B_{GC}$, while the local magnetic field is kept fixed, makes the radial profile of the GMF steeper.
It increases the rate of energy loss of high energy cosmic ray $e^{\pm}$s via synchrotron emission in regions close to the Galactic centre and softens the ICS spectra by up to about $40\%$ as shown in Fig. \ref{fig:ICS_multi_BGC}.
The synchrotron radiation in high latitudes is mostly emitted in the local magnetic field. Since the local magnetic field is kept fixed, the impact of increasing the $B_{GC}$ is smaller in higher latitudes.
In high energies, softening of the ICS spectra is less pronounced. The reason is the following. High energy ICS photons are produced by ICS of high energy comic ray $e^{\pm}$s off of mostly optical and infrared photons. In this regime, the Klein-Nishina scattering cross section must be used. The ICS energy loss rate with $\sigma_{K-N}$ is smaller than the synchrotron energy loss rate which is always proportional to $\sigma_T$.
When these two energy loss rates are very different, a change in one impacts the other one less. Hence, at high energies the effect of increasing the Galactic centre magnetic field on fluxes of ICS is smaller. 

\begin{figure}[htb]
\hspace{-0.5cm}
\includegraphics[scale=0.4]{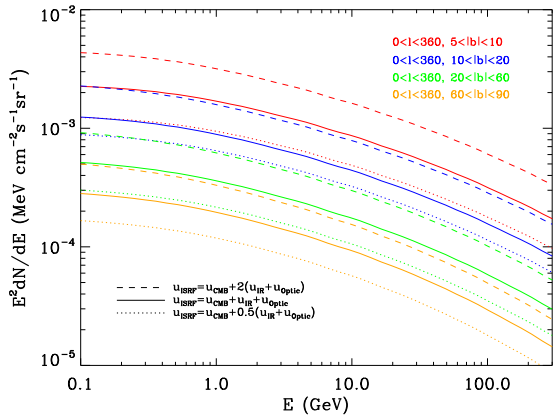}
\caption{Spectrum of ICS in different sky regions for various ISRF energy densities. Fixing the energy density of CMB component, we multiply the energy density of infrared and optical components by $0.5$ (\textit{dotted line}) and $2$ (\textit{dashed line}). The reference ISRF energy density is shown as \textit{solid line}.}
\label{fig:ICS_multi_ISRF}
\end{figure}

The ISRF energy density has a direct impact on the spectrum of ICS as shown in Fig.~\ref{fig:ICS_multi_ISRF}. We fix the CMB energy density and multiply the energy densities of infrared and optical photons by a factor ranging from $0.5$ to $2$. The ICS spectrum becomes harder by about $70\%$ in all sky regions. 

\begin{figure}[htb]
\hspace{-0.5cm}
\includegraphics[scale=0.4]{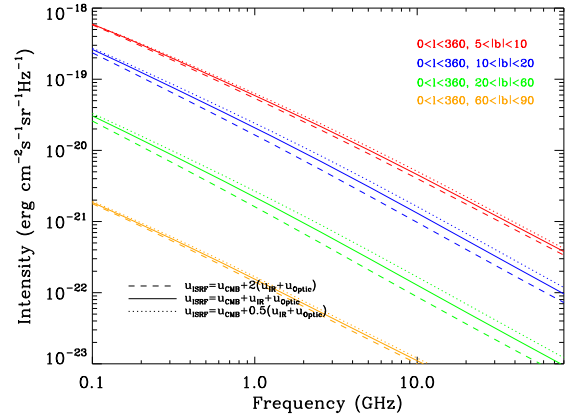}
\caption{Spectrum of synchrotron emission in different sky regions for various ISRF energy densities. Line styles are the same as Fig. \ref{fig:ICS_multi_ISRF}}
\label{fig:Synch_multi_ISRF}
\end{figure}

The increase in energy densities of infrared and optical photons has a reverse impact on the spectra of synchrotron emission and makes them softer.  This effect is more pronounced in intermediate latitudes and it changes the spectra by up to about $50\%$ as shown in Fig.~\ref{fig:Synch_multi_ISRF}. 
In these latitudes the rates of energy loss by synchrotron emission and ICS are close to each other so that  a change in one has a notable effect on the other.
In high latitudes, the rate of energy loss by synchrotron emission is very small because the intensity of the GMF sharply decreases out of the Galactic plane. 
On the other hand in low latitudes, synchrotron radiation is the dominant energy loss mechanism.The reason is that the GMF energy density close to the Galactic plane is much stronger than the ISRF energy density. 
As a result of very different energy loss rates by synchrotron emission and ICS processes in low and high latitudes, varying the ISRF energy density has small effect on synchrotron spectra in those regions.

  
\section{Conclusions}
\label{Conclusion}

In this paper, we addressed an alternative approach to constrain properties of the GMF and the ISRF. 
High energy cosmic ray electrons and positrons that reach us have probed the local interstellar medium thanks to their rapid energy loss. 
Assumptions on local properties of the interstellar medium affect local spectra of cosmic ray $e^{\pm}$s.
High statistics measurements of those spectra by AMS-02 experiment are rather powerful to discriminate between different local budgets of energy density in the GMF and the ISRF and put constraints on them \cite{Maryam}.

The diffuse gamma ray component which is produced by ICS of high energy cosmic ray electrons and positrons off of low energy photons throughout the Galaxy is affected by distribution of the ISRF. 
On the other hand, assumptions on GMF parameters affect ICS spectra by altering the rate of producing gamma ray photons through ICS. 
We quantified those effects in Figs. \ref{fig:ICS_multi_BGC} and \ref{fig:ICS_multi_ISRF}.
High accuracy data on diffuse gamma rays in a wide energy range provided by \textit{Fermi}-LAT and H.E.S.S are able to constrain global properties of both the GMF and the ISRF.
Our forthcoming paper is devoted to analyzing diffuse gamma ray data all over the sky to constrain global properties of these two interstellar medium ingredients.

The spectra and morphology of synchrotron emission are strongly sensitive to global properties of the GMF. 
In fact, using properties of synchrotron emission is the standard approach to determine the distribution of the GMF.
We showed that assumptions on the ISRF energy density have a sizable effect on the spectra of synchrotron emission in intermediate latitudes. This effect should be taken into account in a self consistent analysis of synchrotron emission. 

\vskip 0.2 in
\section*{Acknowledgments}  
\vskip 0.05in
The author warmly thanks Ilias Cholis for valuable discussions since the beginning of this work. I would also like to thank the ICTP for its hospitality during last stages of this work.
 

\bibliography{ISM}
\bibliographystyle{apsrev}

\end{document}